\documentclass[runningheads]{llncs}
\usepackage{graphicx}
\usepackage{float}
\usepackage{tikz}
\usetikzlibrary{shapes.geometric, arrows}
\usepackage{multirow}

\begin{document}
\title{Machine Learning Pipeline for Pulsar Star Dataset
\thanks{\url{https://archive.ics.uci.edu/ml/datasets/HTRU2}}}


\author{Alexander Ylnner Choquenaira Florez\inst{1}\and
Braulio Valentin S\'anchez Vinces\inst{1}\and
Diana Carolina Roca Arroyo\inst{1}\and Josimar Edinson Chire Saire\inst{1} Patr\'icia Batista Franco\inst{1}
%
%
\institute{Institute of Mathematics and Computer Science (ICMC) \\
University of Sao Paulo (USP)\\
Sao Carlos, SP, Brazil}\\
\email{\{alexanderchf, braulio.sanchez, dianaroca, jecs89, patricia.franco\}@usp.br}}

\maketitle              
\begin{abstract}
This work brings together some of the most common machine learning (ML) algorithms, and the objective is to make a comparison at the level of obtained results from a set of unbalanced data. This dataset is composed of almost 17 thousand observations made to astronomical objects to identify pulsars (HTRU2). The methodological proposal based on evaluating the accuracy of these different models on the same database treated with two different strategies for unbalanced data. The results show that in spite of the noise and unbalance of classes present in this type of data, it is possible to apply them on standard ML algorithms and obtain promising accuracy ratios.

\keywords{Pulsar Star \and Machine Learning Application \and Astronomy.}
\end{abstract}

\section{Introduction}
Pulsars are defined as a rare type of stars that, according to research, can emit radio signals which they could be detected from Earth. The Pulsar astronomy area began its existence as a field of study around the year 1967, when Jocelyn Bell discovered by chance a train of pulses spaced regularly with a period of 1.33 seconds in the radio data tables \cite{ref_article9}.

In recent years, scientists in the area have become very interested in this rare type of stars for various reasons. When pulsars rotate, their emission beam travels through the sky. Then, a detectable pattern of broadband radio emission produced when this beam crosses our line of sight. According to research, it can be seen that this pattern is repeated periodically when the pulsars spin rapidly\cite{ref_article9}. 

Machine learning is a method of data analysis to built analytic models. It is a branch of artificial intelligence based on the idea that systems can learn from data, identify patterns, and make decisions with the least human intervention. Also, they can independently modify their behavior based on their own experience. Generally, Machine Learning uses two basics learning paradigm: unsupervised learning and supervised learning. Supervised learning uses labeled data or prior knowledge to estimate the model which represents the data relationship. This work using the supervised learning approach to perform experiments\cite{ref_book1}.

There are many works developed using machine learning techniques for pulsar stars detection \cite{ref_article10,ref_thesis1}, to identify credible by pressing candidates from pulsar surveys using an artificial neural network \cite{ref_article6} and using KNN classifier\cite{ref_article9}, and using image pattern with deep neural nets PICS (Pulsar Image-based Classification System) AI \cite{ref_article7}.

Nowadays, Machine learning tools are used to label pulsar candidates to facilitate rapid analysis automatically. This phenomenon is reduced a binary classification problem. Here the legitimate pulsar examples are a positive minority class, and spurious examples the majority negative class. These examples have all checked by human annotators. 
Each row lists the variables first, and the class label is the final entry. The class labels used are 0 (negative) and 1 (positive). \cite{ref_article11}

This paper, organized as follows: Section 2 describes a literature review about methodologies, processes to explore the dataset and get some insight, Section 3, describes the methodology of the proposal. Section 4, describes the experiments and present the results, and Section 5 states the conclusions.

\section{State of the Art}
A subject of significant research in the field of radio astronomy is the discovery of pulsars. Identifying new pulsar signals in observational radio data can be done either via single pulse searches \cite{ref_article1}, or periodicity searches.
Graphics selection tools such as REAPER \cite{ref_article3}
and JREAPER \cite{ref_article4}
allow the user to project up to several thousand candidates at a time in scatter diagrams. Scoring algorithms such as PEACE \cite{ref_article5}
have also been developed, combining six numerical candidate quality factors into one formula that produces a subjective ranking where pulsars are expected to found close to the top. There ranking method helped to find 47 new pulsars.

With increasingly sophisticated astronomical instruments, the volumes and rates of this type of data are growing exponentially. This fact requires a focus on artificial intelligence (AI) technologies that can perform the automatic identification of the pulsar candidate to extract large sets of astronomical data. \cite{ref_article6} 
Manually created 12 features and trained an artificial neural network of only one hidden layer. In \cite{ref_article2} proposed Straightforward Pulsar Identification using Neural Networks (SPINN) which designs six features and trains a feed-forward single-hiddenlayer artificial neural network for candidate binary classification, their method contributed to 4 new pulsar discoveries. \cite{ref_article7} 
Developed a classification system based on images of pulsars based on the histograms and graphs of the candidates, and training hidden single layer networks, SVM and CNN. They formed an ensemble network combining all the classifiers with logistic regression, although CNN's are powerful in two-dimensional image data, more labeled samples are required for the training of a deep CNN.

In a real scenario, labeling this data is a difficult and expensive process to perform. In the task of selecting candidates for pulsar, the positive samples are very limited because the number of real pulsars discovered is small, and millions of candidates are negative samples.

\section{Methodology}

A diagram of the methodology is:
\tikzstyle{node} = [rectangle, rounded corners, minimum width=2cm, minimum height=1cm,text centered, draw=black, fill=white!30]
\tikzstyle{arrow} = [thick,->,>=stealth]
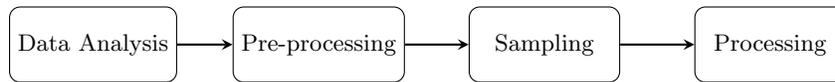
\begin{figure}[H]
  \centerline{
    \begin{tikzpicture}[node distance=3cm]
    \node (dataAnalysis) [node] {Data Analysis};
    \node (preProcessing) [node, right of=dataAnalysis]{Pre-processing};
    \draw [arrow] (dataAnalysis) -- (preProcessing);
    \node (sampling) [node, right of=preProcessing] {Sampling};
    \draw [arrow] (preProcessing) -- (sampling);
    \node (processing) [node, right of=sampling] {Processing};
    \draw [arrow] (sampling) -- (processing);
    \end{tikzpicture}
  }
\caption{ Diagram of the methodology}
\label{fig:1_diagram}
\end{figure}

\subsection{Data Analysis}
Data analysis is performed to see general information about the dataset, the number of items, columns, kind of values (float, string, etc.), missing-values, correlation between features, etc.

\subsection{Pre-processing}
An exploration of dataset is performed to see the distribution of number of items from every target class. This is important because it shows that is an imbalanced dataset. To deal with this, it is performed some techniques of oversampling and undersampling on the original dataset.

\subsection{Sampling}
Sampling is used to choose the training and test set to guarantee the best performance of a machine learning algorithm. It used K-Fold Cross Validation where the dataset is randomly partitioned in k subsamples with the same size. From this k subsamples, one of them used as a testing set, and the other subsamples used as the training set.

\subsection{Processing}
The general process includes a different kind of algorithms. The main objective is to compare the different results. 

The main algorithms are:
\begin{itemize}
    \item Naive Bayes
    \item Logistic Regression
    \item Decision Tree
    \item Perceptron
    \item Multi Layer Perceptron
    \item Support Vector Machine
\end{itemize}
Also, it is applied different kinds of ensembles:
\begin{itemize}
    \item Stacking
    \item Bagging
    \item Random Forest
\end{itemize}

\section{Experiments and Results}
\subsection{Dataset Description}
The dataset used was the HTRU2 that describes a sample of pulsar candidates collected during the Universe Resolution Universe survey. Pulsars are a rare type of neutron star that produces detectable radio emissions here on Earth. The dataset contains 17,898 total examples, of which 1,639 are positive examples, and 16,259 are negative examples. The 16,259 are spurious examples caused by RFI/noise, and 1,639 are real examples of pulsars. Humans annotators checked all examples. In each row, the variables are listed first, and the class label is the final entry. The class labels used are 0 (negative) and 1 (positive), each candidate described by eight continuous variables and a single class variable. The variables contain Mean of the integrated profile; Standard deviation of the integrated profile; Excess kurtosis of the integrated profile; Skewness of the integrated profile; Mean of the DM-SNR curve; Standard deviation of the DMSNR curve; Excess kurtosis of the DM-SNR curve; Skewness of the DM-SNR curve; Class.\cite{ref_article11}

\subsection{Experiments Setup}
The experiments were performed with three variations of the dataset HTRU2 considering sampling to balance unbalanced class and with Feature Selection using Correlation.

\subsubsection{Experiment 1.}
\begin{itemize}
    \item Dataset 1: original dataset with any modification. 
    \item Dataset 2: use of undersampling to decrease the number of sample of the majority class.
    \item Dataset 3: use of oversampling to get synthetic data from minority class to balance the dataset.
\end{itemize}

\begin{table}[]
\centering
\caption{Dataset Dimensions}
\begin{tabular}{|c|c|c|}
\hline
\textbf{Dataset Name} & \textbf{Dimensionality} & \textbf{Distribution of Classes} \\ \hline
Dataset1              & (17898, 9)              & (1639, 16259)                    \\ \hline
Dataset2              & (3278, 9)               & (1639, 1639)                     \\ \hline
Dataset3              & (32518, 9)              & (16259, 16259)                   \\ \hline
\end{tabular}
\end{table}

A PCA visualization with the three main components of the datasets is presented in Fig. \ref{fig:4_dataset}

\begin{figure}[H]
  \centerline{
  \includegraphics[width = 0.3\columnwidth]{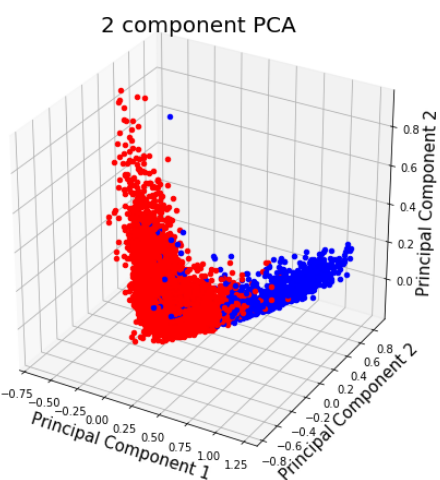}
  \includegraphics[width = 0.3\columnwidth]{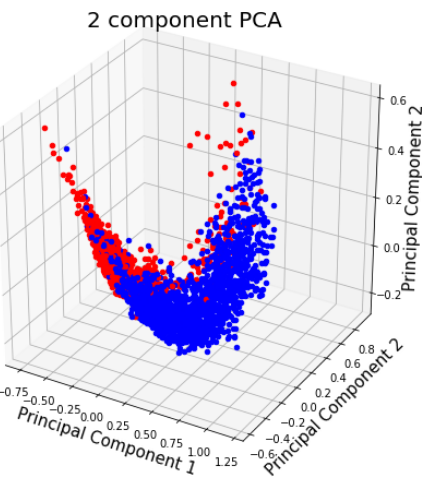}
  \includegraphics[width = 0.3\columnwidth]{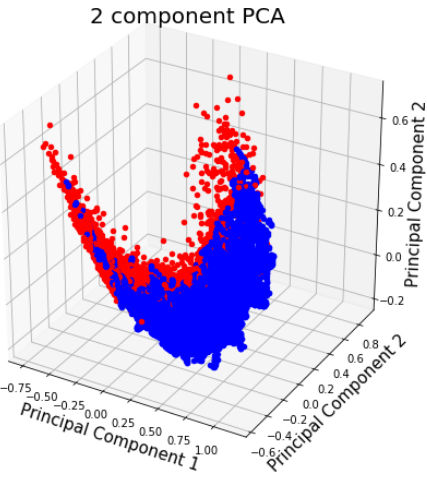}
  }
\caption{ PCA Visualization of Dataset 1, 2, 3}
\label{fig:4_dataset}
\end{figure}

\subsubsection{Experiment 2.}

This experiment is similar to Experiment 1, but considering feature selection, then 6 features are chosen following the correlation proportion between them.
\begin{itemize}
    \item skewness integrated profile
    \item excess kurtosis integrated profile
    \item std dm-snr curve
    \item mean dm-snr curve
    \item skewness dm-snr curve
    \item excess kurtosis dm-snr curve
\end{itemize}{}

A PCA visualization with the three main components of the datasets is presented in Fig. \ref{fig:4_dataset}

\begin{figure}[H]
  \centerline{
  \includegraphics[width = 0.3\columnwidth]{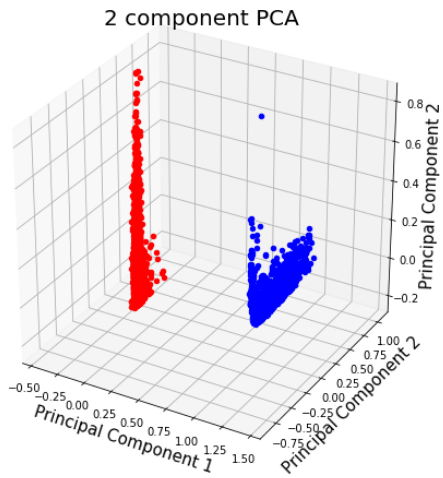}
  \includegraphics[width = 0.3\columnwidth]{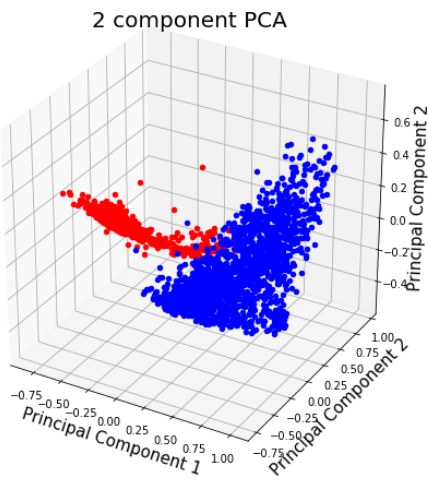}
  \includegraphics[width = 0.3\columnwidth]{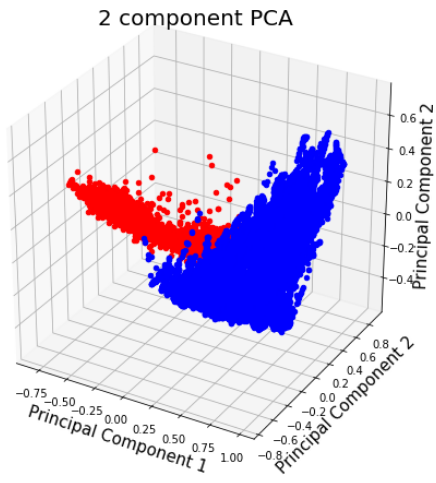}
  }
\caption{ PCA Visualization of Dataset 1, 2, 3}
\label{fig:4_dataset}
\end{figure}

\subsection{Results}

\subsubsection{Experiment 1.}

After to perform the experiments using K-Fold Cross Validation with k = 10 and Accuracy as a measure, the next Tab. \ref{tab:4_tabresults} presents the results.

\begin{table}[H]
\centering
\caption{Results of Experiments}
\begin{tabular}{|c|c|c|c|}
\hline
\textbf{Model}      & \textbf{Dataset 1}       & \textbf{Dataset 2}       & \textbf{Dataset 3}       \\ \hline
Gausssian-NB        & 0.95 (+/- 0.02)          & 0.90 (+/- 0.04)          & 0.90 (+/- 0.01)          \\ \hline
Logistic-Regression & \textbf{0.98 (+/- 0.01)} & 0.93 (+/- 0.03)          & 0.94 (+/- 0.01)          \\ \hline
Decision-Tree       & \textbf{0.98 (+/- 0.01)} & 0.93 (+/- 0.04)          & 0.95 (+/- 0.01)          \\ \hline
Perceptron          & 0.97 (+/- 0.05)          & 0.90 (+/- 0.10)          & 0.86 (+/- 0.19)          \\ \hline
MLPClassifier       & 0.97 (+/- 0.01)          & 0.93 (+/- 0.03)          & 0.92 (+/- 0.01)          \\ \hline
SVC-PolyK           & 0.97 (+/- 0.01)          & 0.92 (+/- 0.04)          & 0.93 (+/- 0.01)          \\ \hline
SVC-RbfK            & 0.97 (+/- 0.01)          & 0.92 (+/- 0.04)          & 0.94 (+/- 0.01)          \\ \hline
SVC-SigK            & 0.97 (+/- 0.01)          & 0.92 (+/- 0.04)          & 0.93 (+/- 0.01)          \\ \hline
Xgboost             & \textbf{0.98 (+/- 0.01)} & \textbf{0.95 (+/- 0.03)} & \textbf{0.95 (+/- 0.00)} \\ \hline
RF                  & 0.97 (+/- 0.01)          & 0.92 (+/- 0.03)          & 0.92 (+/- 0.01)          \\ \hline
Bagging             & \textbf{0.98 (+/- 0.01)} & 0.94 (+/- 0.03)          & 0.97 (+/- 0.01)          \\ \hline
Gradient            & \textbf{0.98 (+/- 0.01)} & 0.94 (+/- 0.04)          & 0.95 (+/- 0.01)          \\ \hline
Stacking            & 0.97  (+/- 0.003)        & 0.91 (+/- 0.02)          & 0.96 (+/- 0.002)         \\ \hline
\end{tabular}
\label{tab:4_tabresults}
\end{table}

Following the results of Tab. \ref{tab:4_tabresults} it is possible to state there is any benefit with undersampling or oversampling to the performance of classifiers even the two algorithms can spend time and resources, but they are not showing the truth about performance classifiers.
The best results are highlight (bold) in Tab. \ref{tab:4_tabresults} in column Dataset 1 (5 values), classifiers got the same result. The same criteria for Dataset 2 and 3.
But, it is necessary to remember the dataset HTRU2 is an unbalanced dataset, and the interest of classification is for class 1. Considering the previous criterion, a report of the experiments is represented in Tab. \ref{tab:4_class_rep} to compare the best with the rest of result classifiers.

\begin{table}[H]
\centering
\caption{Classification Report of Experiments}
\begin{tabular}{|c|c|c|c|c|c|c|c|}
\hline
\multirow{2}{*}{\textbf{Classifier}} & \multirow{2}{*}{\textbf{Metric}} & \multicolumn{2}{c|}{\textbf{Dataset 1}}               & \multicolumn{2}{c|}{\textbf{Dataset 2}}                        & \multicolumn{2}{c|}{\textbf{Dataset 3}}                        \\ \cline{3-8} 
                                     &                                  & 0                         & 1                         & 0                         & 1                                  & 0                         & 1                                  \\ \hline
\multirow{2}{*}{Logistic Regression} & precision                        & 0.97                      & 0.92                      & 0.91                      & \textbf{0.97}                      & 0.91                      & \textbf{0.97}                      \\ \cline{2-8} 
                                     & recall                           & 0.99                      & 0.72                      & 0.98                      & \textbf{0.89}                      & 0.97                      & \textbf{0.91}                      \\ \hline
\multirow{2}{*}{Decision Tree}       & precision                        & 0.98                      & 0.9                       & 0.91                      & \textbf{0.91}                      & 0.94                      & \textbf{0.96}                      \\ \cline{2-8} 
                                     & recall                           & 0.99                      & 0.84                      & 0.92                      & \textbf{0.91}                      & 0.96                      & \textbf{0.94}                      \\ \hline
\multirow{2}{*}{XGboost}             & precision                        & 0.98                      & 0.91                      & 0.92                      & \textbf{0.95}                      & 0.94                      & \textbf{0.97}                      \\ \cline{2-8} 
                                     & recall                           & 0.99                      & 0.83                      & 0.96                      & \textbf{0.92}                      & 0.97                      & \textbf{0.93}                      \\ \hline
\multirow{2}{*}{Bagging}             & precision                        & 0.98                      & 0.9                       & 0.91                      & \textbf{0.95}                      & 0.97                      & \textbf{0.98}                      \\ \cline{2-8} 
                                     & recall                           & 0.99                      & 0.82                      & 0.96                      & \textbf{0.91}                      & 0.98                      & \textbf{0.97}                      \\ \hline
\multirow{2}{*}{Gradient}            & precision                        & 0.98                      & 0.9                       & 0.91                      & \textbf{0.95}                      & 0.93                      & \textbf{0.96}                      \\ \cline{2-8} 
                                     & \multicolumn{1}{l|}{recall}      & \multicolumn{1}{l|}{0.99} & \multicolumn{1}{l|}{0.79} & \multicolumn{1}{l|}{0.96} & \multicolumn{1}{l|}{\textbf{0.91}} & \multicolumn{1}{l|}{0.96} & \multicolumn{1}{l|}{\textbf{0.93}} \\ \hline
\end{tabular}
\label{tab:4_class_rep}
\end{table}

Then, following the classification report is confirmed the previous idea, classifiers got best results with the Dataset 1 because of the majority class 0 but considering the application of identifying pulsar star is most important to identify class 1. According to results is confirmed a sampling for an unbalanced dataset is necessary for spite of the initial results was the best. Now, considering sampled datasets (Dataset 2, 3) is easy to notice the benefit of sampling and oversample, the minority class got the best results for class 1 (Dataset 3).

\subsubsection{Experiment 2.}

The Tab. \ref{tab:4_results2} presents results of Experiment 2, a similar situation found here. Classifiers got best results with Dataset 1. But there is a small increment in accuracy metric because of the Feature Selection.

\begin{table}[H]
\centering
\caption{Results Experiment 2}
\begin{tabular}{|c|c|c|c|}
\hline
\textbf{Model}      & \textbf{Dataset 1}       & \textbf{Dataset 2} & \textbf{Dataset 3} \\ \hline
Gausssian-NB        & 0.94 (+/- 0.02)          & 0.90 (+/- 0.05)    & 0.90 (+/- 0.01)    \\ \hline
Logistic-Regression & 0.97 (+/- 0.01)          & 0.92 (+/- 0.05)    & 0.94 (+/- 0.01)    \\ \hline
Decision-Tree       & \textbf{0.98 (+/- 0.01)} & 0.94 (+/- 0.04)    & 0.95 (+/- 0.01)    \\ \hline
Perceptron          & 0.97 (+/- 0.01)          & 0.93 (+/- 0.05)    & 0.90 (+/- 0.12)    \\ \hline
MLPClassifier       & 0.96 (+/- 0.01)          & 0.92 (+/- 0.04)    & 0.92 (+/- 0.01)    \\ \hline
SVC-PolyK           & 0.97 (+/- 0.01)          & 0.90 (+/- 0.05)    & 0.93 (+/- 0.01)    \\ \hline
SVC-RbfK            & \textbf{0.98 (+/- 0.01)} & 0.92 (+/- 0.04)    & 0.94 (+/- 0.01)    \\ \hline
SVC-SigK            & 0.97 (+/- 0.01)          & 0.90 (+/- 0.05)    & 0.93 (+/- 0.01)    \\ \hline
Xgboost             & \textbf{0.98 (+/- 0.01)} & 0.95 (+/- 0.04)    & 0.95 (+/- 0.01)    \\ \hline
RF                  & \textbf{0.98 (+/- 0.01)} & 0.94 (+/- 0.04)    & 0.94 (+/- 0.01)    \\ \hline
Bagging             & \textbf{0.98 (+/- 0.01)} & 0.94 (+/- 0.04)    & 0.97 (+/- 0.01)    \\ \hline
Gradient            & \textbf{0.98 (+/- 0.01)} & 0.94 (+/- 0.04)    & 0.95 (+/- 0.01)    \\ \hline
Stacking            & 0.97 (+/- 0.00)          & 0.92 (+/- 0.02)    & 0.95 (+/- 0.00)    \\ \hline
\end{tabular}
\label{tab:4_results2}
\end{table}

\begin{table}[H]
\centering
\caption{Classification Report on Experiment 2}
\begin{tabular}{|c|c|c|c|c|c|c|c|}
\hline
\multirow{2}{*}{\textbf{Classifier}} & \multirow{2}{*}{\textbf{Metric}} & \multicolumn{2}{c|}{\textbf{Dataset 1}} & \multicolumn{2}{c|}{\textbf{Dataset 2}} & \multicolumn{2}{c|}{\textbf{Dataset 3}} \\ \cline{3-8} 
                                     &                                  & 0                  & 1                  & 0              & 1                      & 0              & 1                      \\ \hline
\multirow{2}{*}{Decision-Tree}       & precision                        & 0.98               & 0.87               & 0.92           & \textbf{0.92}          & 0.94           & \textbf{0.96}          \\ \cline{2-8} 
                                     & recall                           & 0.99               & 0.84               & 0.93           & \textbf{0.92}          & 0.96           & \textbf{0.94}          \\ \hline
\multirow{2}{*}{SVC-RbfK}            & precision                        & 0.97               & 0.96               & 0.89           & \textbf{0.98}          & 0.9            & \textbf{0.98}          \\ \cline{2-8} 
                                     & recall                           & 1                  & 0.71               & 0.98           & \textbf{0.88}          & 0.98           & \textbf{0.9}           \\ \hline
\multirow{2}{*}{Xgboost}             & precision                        & 0.98               & 0.9                & 0.92           & \textbf{0.98}          & 0.93           & \textbf{0.97}          \\ \cline{2-8} 
                                     & recall                           & 0.99               & 0.83               & 0.98           & \textbf{0.91}          & 0.97           & \textbf{0.93}          \\ \hline
\multirow{2}{*}{RF}                  & precision                        & 0.97               & 0.92               & 0.91           & \textbf{0.99}          & 0.91           & \textbf{0.97}          \\ \cline{2-8} 
                                     & recall                           & 0.99               & 0.75               & 0.99           & \textbf{0.9}           & 0.97           & \textbf{0.91}          \\ \hline
\multirow{2}{*}{Bagging}             & precision                        & 0.98               & 0.89               & 0.91           & \textbf{0.96}          & 0.96           & \textbf{0.97}          \\ \cline{2-8} 
                                     & recall                           & 0.99               & 0.81               & 0.96           & \textbf{0.9}           & 0.97           & \textbf{0.96}          \\ \hline
\multirow{2}{*}{Gradient}            & precision                        & 0.98               & 0.9                & 0.92           & \textbf{0.97}          & 0.94           & \textbf{0.96}          \\ \cline{2-8} 
                                     & recall                           & 0.99               & 0.79               & 0.97           & \textbf{0.91}          & 0.96           & \textbf{0.94}          \\ \hline
\end{tabular}
\label{tab:4_classrep_2}
\end{table}

Considering the previous situation of Experiment 1, it was expected to have a better metric with class 1 for Dataset 2, 3. 

\section{Conclusions}

\begin{itemize}
    \item Pulsar Start Dataset HTRU2 is an unbalanced dataset, this feature influence in the pipeline of Machine Learning to use in the experiments.
    \item Sampling is necessary to balance minority class and have appropriate training to reduce bias.
    \item Classifiers can show the best results, but it is necessary to analyze the result for class and choose the importance class for the application.
\end{itemize}{}

\bibliographystyle{splncs04}
\bibliography{biblio}

\begin{thebibliography}{10}
\providecommand{\url}[1]{\texttt{#1}}
\providecommand{\urlprefix}{URL }
\providecommand{\doi}[1]{https://doi.org/#1}

\bibitem{ref_book1}
Alpaydin, E.: Introduction to machine learning. Cambridge, Mass. : MIT Press,
  c2010., 2 edn. (2010)

\bibitem{ref_article1}
Cordes, J.M., McLaughlin, M.A.: Searches for fast radio transients. The
  Astrophysical Journal  \textbf{596}(2),  1142--1154 (oct 2003).
  \doi{10.1086/378231}, \url{https://doi.org/10.1086\%2F378231}

\bibitem{ref_article6}
Eatough, R.P., Molkenthin, N., Kramer, M., Noutsos, A., Keith, M.J., Stappers,
  B.W., Lyne, A.G.: {Selection of radio pulsar candidates using artificial
  neural networks}. Monthly Notices of the Royal Astronomical Society
  \textbf{407}(4),  2443--2450 (07 2010).
  \doi{10.1111/j.1365-2966.2010.17082.x},
  \url{https://doi.org/10.1111/j.1365-2966.2010.17082.x}

\bibitem{ref_article3}
Faulkner, A.J., Stairs, I.H., Kramer, M., Lyne, A.G., Hobbs, G., Possenti, A.,
  Lorimer, D.R., Manchester, R.N., McLaughlin, M.A., D'Amico, N., Camilo, F.,
  Burgay, M.: {The Parkes Multibeam Pulsar Survey – V. Finding binary and
  millisecond pulsars}. Monthly Notices of the Royal Astronomical Society
  \textbf{355}(1),  147--158 (11 2004). \doi{10.1111/j.1365-2966.2004.08310.x},
  \url{https://doi.org/10.1111/j.1365-2966.2004.08310.x}

\bibitem{ref_article4}
Keith, M.J., Eatough, R.P., Lyne, A.G., Kramer, M., Possenti, A., Camilo, F.,
  Manchester, R.N.: {Discovery of 28 pulsars using new techniques for sorting
  pulsar candidates}. Monthly Notices of the Royal Astronomical Society
  \textbf{395}(2),  837--846 (04 2009). \doi{10.1111/j.1365-2966.2009.14543.x},
  \url{https://doi.org/10.1111/j.1365-2966.2009.14543.x}

\bibitem{ref_article5}
Lee, K.J., Stovall, K., Jenet, F.A., Martinez, J., Dartez, L.P., Mata, A.,
  Lunsford, G., Cohen, S., Biwer, C.M., Rohr, M., Flanigan, J., Walker, A.,
  Banaszak, S., Allen, B., Barr, E.D., Bhat, N.D.R., Bogdanov, S., Brazier, A.,
  Camilo, F., Champion, D.J., Chatterjee, S., Cordes, J., Crawford, F., Deneva,
  J., Desvignes, G., Ferdman, R.D., Freire, P., Hessels, J.W.T., Karuppusamy,
  R., Kaspi, V.M., Knispel, B., Kramer, M., Lazarus, P., Lynch, R., Lyne, A.,
  McLaughlin, M., Ransom, S., Scholz, P., Siemens, X., Spitler, L., Stairs, I.,
  Tan, M., van Leeuwen, J., Zhu, W.W.: {peace: pulsar evaluation algorithm for
  candidate extraction – a software package for post-analysis processing of
  pulsar survey candidates}. Monthly Notices of the Royal Astronomical Society
  \textbf{433}(1),  688--694 (05 2013). \doi{10.1093/mnras/stt758},
  \url{https://doi.org/10.1093/mnras/stt758}

\bibitem{ref_article11}
Lyon, R.J., Stappers, B.W., Cooper, S., Brooke, J.M., Knowles, J.D.: {Fifty
  Years of Pulsar Candidate Selection: From simple filters to a new principled
  real-time classification approach}. Mon. Not. Roy. Astron. Soc.
  \textbf{459}(1),  1104--1123 (2016). \doi{10.1093/mnras/stw656}

\bibitem{ref_article10}
McFadden, R., Karastergiou, A., Roberts, S.: Machine learning for pulsar
  detection. Proceedings of the International Astronomical Union
  \textbf{13}(S337),  372–373 (2017). \doi{10.1017/S1743921317009000}

\bibitem{ref_article9}
Mohamed, T.M.: Pulsar selection using fuzzy knn classifier. Future Computing
  and Informatics Journal  \textbf{3}(1),  1 -- 6 (2018).
  \doi{https://doi.org/10.1016/j.fcij.2017.11.001},
  \url{http://www.sciencedirect.com/science/article/pii/S2314728817300776}

\bibitem{ref_article2}
Morello, V., Barr, E.D., Bailes, M., Flynn, C.M., Keane, E.F., van Straten, W.:
  {SPINN: a straightforward machine learning solution to the pulsar candidate
  selection problem}. Monthly Notices of the Royal Astronomical Society
  \textbf{443}(2),  1651--1662 (07 2014). \doi{10.1093/mnras/stu1188},
  \url{https://doi.org/10.1093/mnras/stu1188}

\bibitem{ref_thesis1}
Morello, V.: {Discovering Pulsars with Machine Learning}. Ph.D. thesis, Faculty
  of Science, Engineering and Technology Swinburne University (May 2016)

\bibitem{ref_article7}
Zhu, W.W., Berndsen, A., Madsen, E.C., Tan, M., Stairs, I.H., Brazier, A.,
  Lazarus, P., Lynch, R., Scholz, P., Stovall, K., Ransom, S.M., Banaszak, S.,
  Biwer, C.M., Cohen, S., Dartez, L.P., Flanigan, J., Lunsford, G., Martinez,
  J.G., Mata, A., Rohr, M., Walker, A., Allen, B., Bhat, N.D.R., Bogdanov, S.,
  Camilo, F., Chatterjee, S., Cordes, J.M., Crawford, F., Deneva, J.S.,
  Desvignes, G., Ferdman, R.D., Freire, P.C.C., Hessels, J.W.T., Jenet, F.A.,
  Kaplan, D.L., Kaspi, V.M., Knispel, B., Lee, K.J., van Leeuwen, J., Lyne,
  A.G., McLaughlin, M.A., Siemens, X., Spitler, L.G., Venkataraman, A.:
  {SEARCHING} {FOR} {PULSARS} {USING} {IMAGE} {PATTERN} {RECOGNITION}. The
  Astrophysical Journal  \textbf{781}(2), ~117 (jan 2014).
  \doi{10.10088/0004-637x/781/2/117},
  \url{https://doi.org/10.1088\%2F0004-637x\%2F781\%2F2\%2F117}

\end{thebibliography}

\end{document}